# SYNERGETICS IN MULTIPLE EXCITON GENERATION EFFECT IN QUANTUM DOTS

Turaeva N.N., Oksengendler B.L., Uralov I.

Institute of polymer chemistry and physics

100128 Kadiri str.7B, Tashkent, Uzbekistan

e-mail: nturaeva@hotmail.com

**Abstract**. We present detailed analysis of the non-Poissonian population of excitons produced by MEG effect in quantum dots on the base of statistic theory of MEG and synergetic approach for chemical reactions. From the analysis we can conclude that a non-Poissonian distribution of exciton population is evidence of non-linear and non-equilibrium character of the process of multiple generation of excitons in quantum dots at a single photon absorption.

There are two conceptual principles in nanostructures, which distinguish nanostructures from their bulk counterparts: a quantum confinement of carriers and a significant ratio of surface and bulk atoms. Quantum confinement of carriers in quantum dots leads to the two consequences: quantization of energy states, tunable by quantum dots size and large exciton and biexciton energies. As a result of these two effects of quantum confinement, multiple exciton generation (MEG) effect occurs in quantum dots at absorption of a single photon. This effect was predicted [1] and experimentally observed by many researchers [2]. It has been revealed the principal role of surface of quantum dots in MEG effect [3]. Up to now there are several hypothesis of the effect [4-6], in a frame of which it is difficult to explain a non-Poissonian character of multiplicity of excitons generating by single photon in quantum dot [7]. In this work a non-Poissonian character of excitons multiplicity in PbSe quantum dots will be discussed in a frame of statistic theory of MEG [8,9] and via a master-equation for distribution function of exciton multiplicity basing on synergetics approaches to chemical reactions [10-12].

*Non-Poissonian character of exciton multiplicity on the base of statistical theory.* In a frame of the statistic theory of MEG [8], which developed on the base of Fermi approach to simultaneous generation of mesons at nucleon-nucleon collisions, it can be possible to calculate not only average multiplicity, but the distribution of fluctuations of exciton multiplicity. The distribution of exciton multiplicity is defined as

$$P(n) = \frac{S(n)}{\sum_n S(n)}. \qquad (1)$$

Here n is the number of electrons and holes, S(n) is the statistical weight of generation of n particles in quantum dot of size R at absorption of a single photon ($h\nu$) which can be calculated from the expression [8,9]

$$S(n) = \frac{m^{3n/2} V^n}{2^{3n/2} \pi^{3n/2} \hbar^{3n}} \frac{\left(h\nu - \frac{n}{2}\tilde{E}_g\right)^{\frac{3n}{2}-1}}{\left(\frac{3n}{2}-1\right)!}.$$

(2)

Here m is the mass of electrons (holes), V is the volume of quantum dot, $\tilde{E}_g$ is the effective energy of energy gap with taking into account exciton energy. The average number of excitons $\langle N_{exc} \rangle$ generated by a single photon is defined on the base of following formula

$$\bar{n} = 2\langle N_{exc} \rangle = \frac{\sum_n n S(n)}{\sum_n S(n)} = \sum_n n P(n). \quad (3)$$

Taking into account the data on PbSe quantum dot size and photon energy [4], at which the MEG effect was exhibited, it can be possible to plot the function of exciton distribution P(n) (fig.1). It is easily to show by means of the statistic theory of MEG effect that the principal criterion of Poissonian distribution, that is, $\overline{n^2} = (\bar{n})^2 + \bar{n}$, is not satisfied. For example, for a PbSe quantum dot of R=3.9 nm we have $\bar{n} = 4.2$; $(\bar{n})^2 = 17.64$; $\overline{n^2} = 18.4$ at absorption of photon with $h\nu = 3.63 E_g$ (curve (a) of fig.1) and $\bar{n} = 5.7$; $(\bar{n})^2 = 32.49$; $\overline{n^2} = 33.46$ at absorption of photon with $h\nu = 4.9 E_g$ (curve (b) of fig.1), that is, $\overline{n^2} \neq (\bar{n})^2 + \bar{n}$. It means that the exciton multiplicity distribution function is of non-Poissonian character. Moreover, the deviation from the Poissonian distribution increases with the photon energy $\left(\frac{d\left(\overline{n^2} - (\bar{n})^2 + \bar{n}\right)}{d(h\nu)} > 0\right)$.

*Master-equation of MEG in quantum dots.* In general case, for chemical reactions occurring in systems, where fluctuations are not considered negligible small, it is necessary to compose master-equation for distribution of probability [10-12]. Analyzing the monomolecular reaction of a type as A↔X↔E on the base of master-equation it can be deduced the Poissonian distribution with average value of X. It is known [12] that at considering such reactions as A+X→2X the corresponding kinetic equations for markovian processes become non-linear and this peculiarity leads to the non-Poissonian distribution functions [10-12]. This result was proved by Nicolis and Prigogine and caused a great scientific interest. Let us consider a master-equation for chemical reactions of the process of multiple exciton generation in quantum dots at absorption of a single photon and analyze its solution. For simplicity we limit by the case of two excitons generation, that is

$$A + X \underset{k_{-1}}{\overset{k_1}{\Longleftrightarrow}} 2X$$

$$X \underset{k_{-2}}{\overset{k_2}{\Longleftrightarrow}} A. \qquad (4)$$

Here A is the concentration of valence electrons, excitons is denoted by X, $k$ and $k_-$ are the constants of direct and reverse chemical reactions. The first reaction corresponds to the generation of two excitons due to the interaction of a single exciton with the valence electron and their relaxation into the single exciton state and the second reaction corresponds to the processes of recombination and generation of a single exciton. Denoting the number of excitons by N we can define a master equation for the distribution of multiplicity fluctuations P(N,t) taking into account two types of transition - the exciton generation ($N \to N+1, N-1 \to N$) and the exciton annihilation ($N+1 \to N$, $N \to N-1$) for the two reactions above:

$$\dot{P}(N,t) = P(N-1,t)W(N, N-1) + P(N+1,t) \times \\ \times W(N, N+1) - P(N,t) \times \\ \times [W(N+1, N) + W(N-1, N)], \qquad (5)$$

$$W(N, N-1) = \left(k_1 A \frac{N-1}{V} + k_{-2} A\right)V$$

$$W(N, N+1) = \left(k_{-1} \frac{(N+1)N}{V^2} + k_2 \frac{N+1}{V}\right)V. \qquad (6)$$

Here W is the transition probability, V is the volume of quantum dot. The stationary solution of the master-equation (5) is defined as

$$P(N) = P(0) \prod_{n=0}^{N-1} \frac{W((n+1,n)}{W(n,n+1)} = P(0) \times$$

$$\times \prod_{n=0}^{N-1} \frac{k_1 AnV + k_{-2}AV^2}{k_{-1}(n+1)n + k_2 V(n+1)}. \qquad (7)$$

Denoting the probabilities of two transitions by the following expressions

$$W(N, N-1) = \omega_+(N) \\ W(N, N+1) = \omega_-(N) \qquad (8)$$

and basing on the condition of extremes of P(N) [10]

$$\frac{\omega_+(N_0+1)}{\omega_-(N_0)} = 1, \qquad (9)$$

we can find the extremum solutions $N_0$ of the equation (7):

$$N_0^{1,2} = \frac{k_{-1} + k_2 V - k_1 AV \pm}{2k_{-1}}$$

$$\frac{\pm \sqrt{(k_1 AV - k_{-1} - k_2 V)^2 - 4k_{-1}(k_{-2}AV^2 + k_2 V)}}{2k_{-1}}.$$

(10)

From (10) it is obvious that the positive value of $N_0$ are derived at satisfaction of the condition

$$k_{-1} + k_2 V > k_1 AV. \qquad (11)$$

The probability $P(N_0)$ is maximal at the following conditions [10]:

$$\begin{cases} \dfrac{k_1 A(N_o - 1)V + k_{-2}AV^2}{k_{-1}N_0(N_0-1) + k_2 N_0 V} > 1 \\ \dfrac{k_1 A(N_o + 1)V + k_{-2}AV^2}{k_{-1}(N_0+2)(N_0-1) + k_2(N_0+2)V} < 1. \end{cases} \qquad (12)$$

Solving the inequalities (12), we can receive the condition of $N_0 > -\dfrac{k_{-1} + k_2 V - k_1 AV}{k_{-1}}$.

This inequality is always satisfied due to the (11). Hence the extremum solutions are maximum values. Thus the graph of P(N) represents the curve with two maxima. For our case, when the process of MEG occurs almost instantaneously (<50-200fs) compared to the process of their relaxation into a single exciton state (~300ps) [4], which corresponds to the great ratio of $k_1/k_{-1}$, it is easily to show on the base of (7) and (9) that there is just one maximum in distribution function of P(N), that is

$$N_0 = \frac{k_2 - k_{-2}AV}{k_1 A - k_2}. \qquad (13)$$

In general case, the distribution function P(N) (7) is of non-Poissonian character. Using the principle of detailed balance for both chemical reactions and presenting the transition probabilities (6) as the sum of the probabilities, related to two reactions we can derive the Poissonian distribution function. For the non-equilibrium processes, the detailed-balance principle is not valid and we deduce the distribution function differed from the Poissonian one.

Thus, from the detailed analysis basing on synergetics approaches and statistical theory of MEG effect we can conclude that the process of multiple exciton generation in quantum dots is of non-linear and non-equilibrium character and therefore the population of excitons exhibit a non-Poissonian distribution.